\documentclass[twocolumn,letterpaper,amsmath,amssymb]{revtex4} 
\usepackage{graphicx}
\usepackage{dcolumn}
\usepackage{bm}
\usepackage{graphicx}
\usepackage{slashbox}
\usepackage[dvips]{epsfig}
\usepackage{float}

\begin{document}
  
\title{Model of random packings of different size balls}

\author{Maximilien Danisch$^{1}$, Yuliang Jin$^{2}$, Hernan A. Makse$^2$}

\affiliation{$^1$ Physics Departement, Ecole Normale Superieure de Cachan, 61 avenue du President Wilson, 94235 Cachan, France.\\
  $^2$ Levich Institute and Physics Department, City College of New York, New York, NY 10031, US}

\begin{abstract}
  We develop a model to describe the properties of random assemblies
  of polydisperse hard spheres. We show that the key features to
  describe the system are {\it (i)} the dependence between the free
  volume of a sphere and the various coordination numbers between the
  species, and {\it (ii)} the dependence of the coordination numbers
  with the concentration of species; quantities that are calculated
  analytically. The model predicts the density of random close packing
  and random loose packing of polydisperse systems for a given
  distribution of ball size and describes packings for any
  interparticle friction coefficient.  The formalism allows to
  determine the optimal packing over different distributions and may
  help to treat packing problems of non-spherical particles which are
  notoriously difficult to solve.
\end{abstract}


\maketitle

Understanding the basic properties of spheres packings is a major
challenge since this problem may provide valuable knowledge regarding
low temperature phases in condensed matter physics
\cite{coniglio}. The canonical example is perhaps the monodisperse
sphere packing problem. It has been mathematically proven that the
optimum way to arrange monodisperse spheres is the face-centered cubic
lattice; a problem that has been solved recently by Hales, $\sim$ 400
years after the famous Kepler conjecture on the issue. On the other
hand, it is commonly observed that packings arrange in a random
fashion at a lower density state called random close packing or RCP
\cite{bernal}. Furthermore, packings are mechanically stable up to an
even lower limit called random loose packing, RLP.

In parallel with the large literature dealing with monodisperse sphere
packings, a large body of experimental, theoretical and numerical work
has been devoted to the analysis of polydisperse systems; the interest
arising due to the simple fact that polydispersivity is omnipresent in
most realistic systems and industrial applications
\cite{doods,zamponi}.
While previous approaches have focused on frictionless packings, an
integrated analytical approach that brings together different
observations for all packings from RLP to RCP and for any friction or
coordination number is still lacking. Based on our previous
statistical mechanics approach \cite{swm}, here we build such a
framework.

We show that the key aspect to solve this problem is the dependence of
the various coordination numbers between the different species and the
concentration of the species. This is calculated here and shown to
agree well with computer simulations. This result is then incorporated
into a statistical theory of volume fluctuations as in \cite{swm}
which calculates the free volume of a particle in terms of the
coordination number. The main result is the prediction of the RLP and
RCP limiting densities for a given distribution of ball sizes as well
as the prediction of densities for any packing in between those
limits. The formalism allows for a determination of the best packing
fraction in terms of different distribution of ball sizes with
specified constraints, as we show with a simple example. We discuss
possible generalization of the method to solve more difficult problems
like the phase behavior of systems of non-spherical particles like
rods or spherocylinders in any dimensions; problems of long-standing
history in condensed matter \cite{onsager}.

Recent theoretical advances \cite{swm} allow for the prediction of the density of RCP and RLP for equal-size ball packings using a relation between the average volume and the geometrical coordination number. 
Following this approach, we here describe long-range
spatial correlations through a mean-field background term. This approximation
makes the problem amenable to analytic calculations, and is shown to describe well
our simulation results. An explicit inclusion of such correlations is possible in
our framework, but severely complicates any solution attempts. Thus, we believe that
the present approach is accurate enough for many important properties, such as the volume fraction calculation.



The above theoretical framework will guide the present formalism for polydisperse systems. We
first treat the case of binary mixtures of hard spheres of radius
$R_1$ and $R_2>R_1$ in 3d and then generalize the problem to any
distribution and dimension.

{\bf Calculation of $z_{ij}$.---} The key quantity to calculate is
$z_{ij}$, denoting the mean number of contacts of a ball of radius
$R_i$ with a ball of radius $R_j$, versus the concentration of one of
the species.  We need a formula for $z_{ij}$ as a function of
$(z,x,\frac{R_2}{R_1})$, the later being the size ratio, $x$ is the
fraction of small balls in the packing $x\equiv N_1/(N_1+N_2)$ with
$N_i$ the number of $i$ balls and $z$ is the global geometrical
coordination number averaged over all the particles: $z \equiv x z_1 +
(1-x) z_2$ where $z_i$ is the average coordination of each species.

\begin{figure}[t]
\hbox{
 (a)\resizebox{0.22\textwidth}{!}{\includegraphics{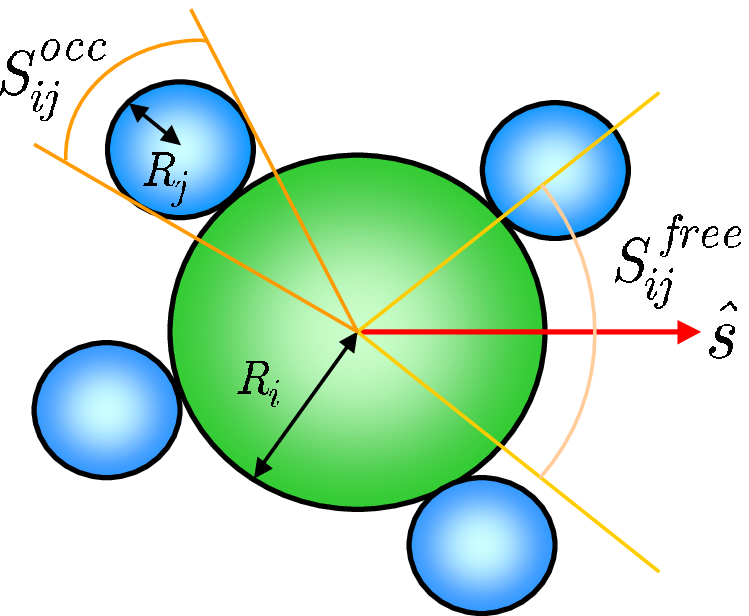}}
 (b)\resizebox{0.22\textwidth}{!}{\includegraphics{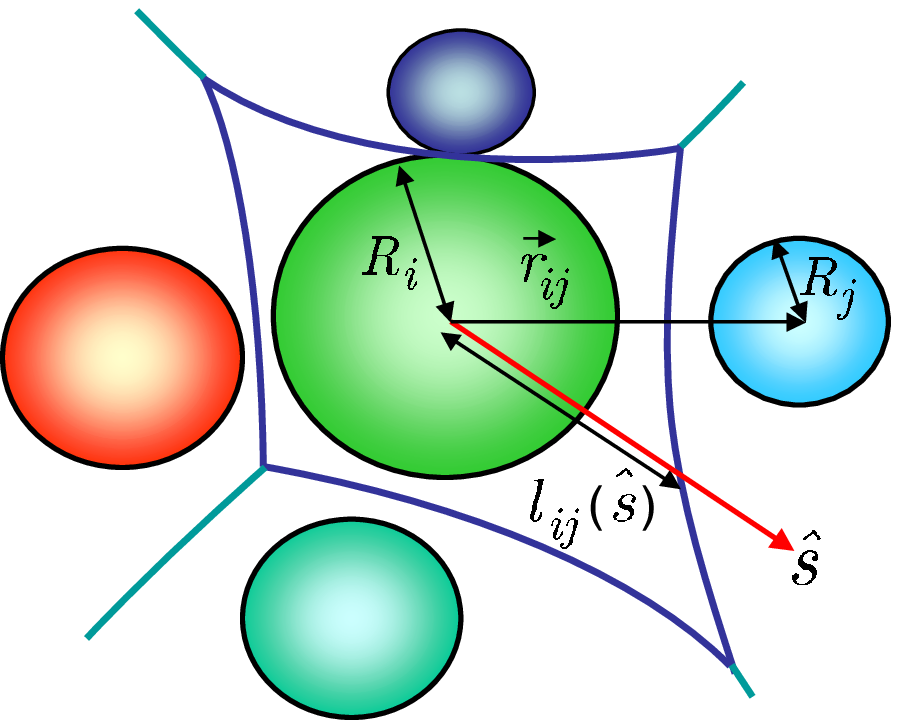}}}
\caption{(a) Occupied surface and free surface.  (b) Voronoi cell for
  polydisperse balls. Plots are in 2d for easier visualization.}
    \label{voro}
\end{figure}

The coordinations are determined by three equations:
\begin{equation}
z_{i}=z_{i1}+z_{i2}, \,\,(i=1,2), \,\,\,\,\, \,\,\,\,\,\,\,
x z_{12}=(1-x) z_{21}.
\label{z1221}
\end{equation}
We assume that these coordinations are inversely proportional to the
average solid angle extended by contacting balls $R_1$ and $R_2$.
The average solid angles are denoted $\langle S^{occ}_i\rangle$ and
are calculated in terms of the solid angle that a ball $R_j$ occupy on
a ball $R_i$ according to $\langle S_i^{occ} \rangle =x
S_{i1}^{occ}+(1-x)S_{i2}^{occ}$ with (see Fig. \ref{voro}a)
\begin{equation}
  S^{occ}_{ij}/2\pi \equiv
  \int^{\arcsin(\frac{R_j}{R_i+R_j})}_{0}\sin \theta \, d\theta =
  1-\sqrt{1-(\frac{R_j}{R_i+R_j})^2}.\nonumber
\label{occ}
\end{equation}

Thus, $\langle S_i^{occ}\rangle$ represents the mean occupied surface
on a $i$ ball weighted by the concentrations $x$ and $(1-x)$. This
represents an approximation since the real weights are $z_{i1}/z_i$
and $z_{i2}/z_i$, respectively. Then, $z_i\propto 1/ \langle
S_{i}^{occ} \rangle$ leading to the following normalizations:

\begin{equation}
  z_1=\frac{z}{x + (1-x) \frac{\langle S^{occ}_1\rangle}{\langle S^{occ}_2\rangle}}, \,\,\,\,\,\,\,z_2 =\frac{z}{x \frac{\langle S^{occ}_2\rangle}{\langle S^{occ}_1\rangle} +(1-x) }.
\label{z2}
\end{equation}

Thus, the system of Eqs. (\ref{z1221}) is reduced to a system of three
equations for four unknowns $z_{ij}$. To close the system we assume
proportional laws and deduce $z_{ij}$ from $z_i$ by considering that
$z_{ij}$ is proportional to the number of contacts of the $i$ balls times
the number of contacts of the $j$ balls:
\begin{eqnarray}
  z_{11}=A(z_1x)(z_1x),& \,\,\,\,\,\,& z_{12}=A (z_1x) z_2(1-x), \\
  z_{21}=Bz_2(1-x)(z_1x),& \,\,\,\,\,\, &
  z_{22}=B z_2(1-x) z_2(1-x).  \nonumber
\end{eqnarray}
Using the first equation in (\ref{z1221}) we find the constants $A$
and $B$, leading to the solution:
\begin{eqnarray}
  z_{11}=\frac{z_1^2 x}{z}, & \,\,\,\,\,\,& z_{12}=\frac{z_1 z_2 (1-x)}{z},
\nonumber
\\
z_{21}=\frac{z_1 z_2 x}{z}, & \,\,\,\,\,\,& z_{22}=\frac{z_2^2 (1-x)}{z}.
\label{zs}
\end{eqnarray}

\begin{figure}[t]
\hbox{
(a)        \includegraphics[width=0.21\textwidth]{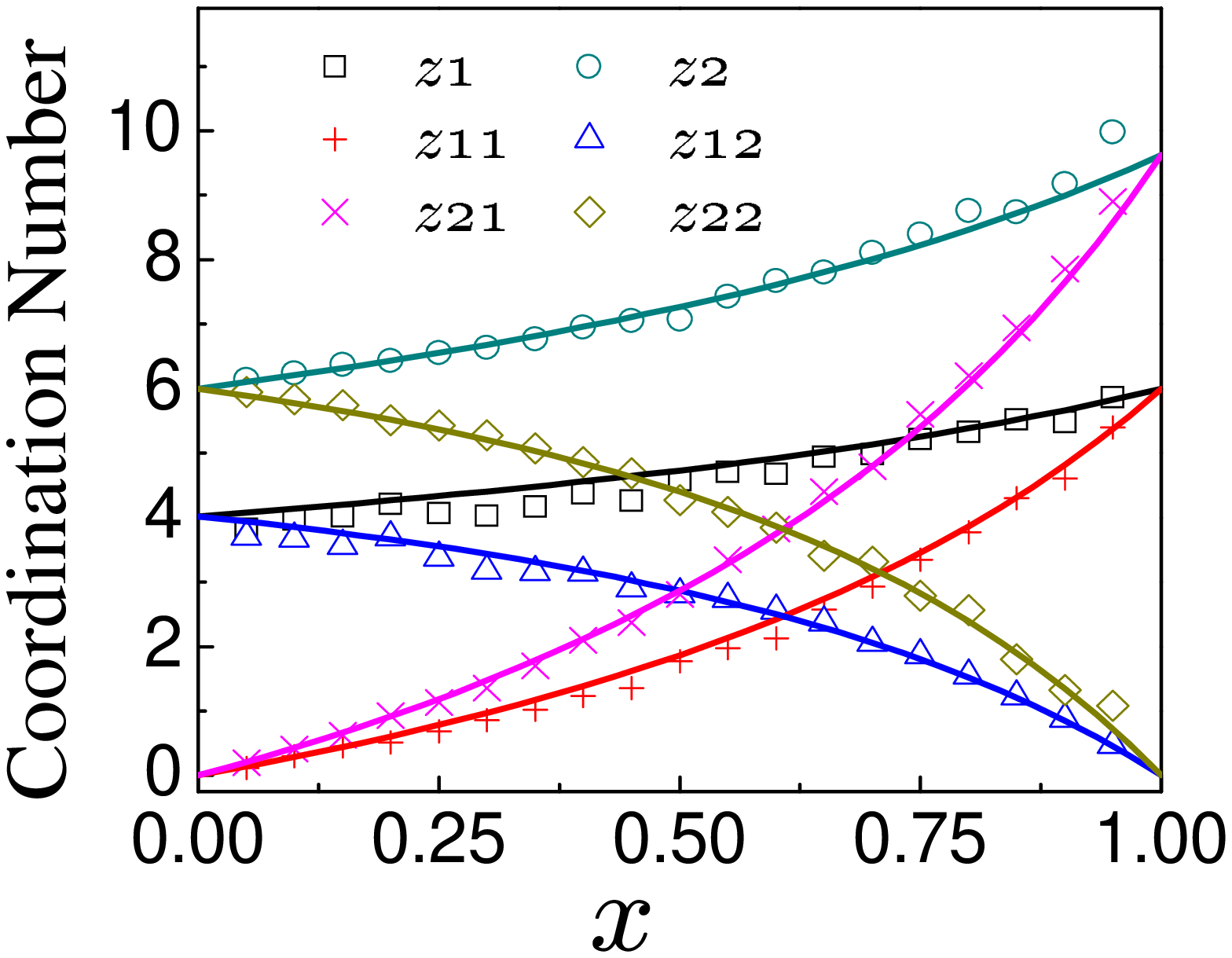} (b)
\includegraphics[width=0.21\textwidth]{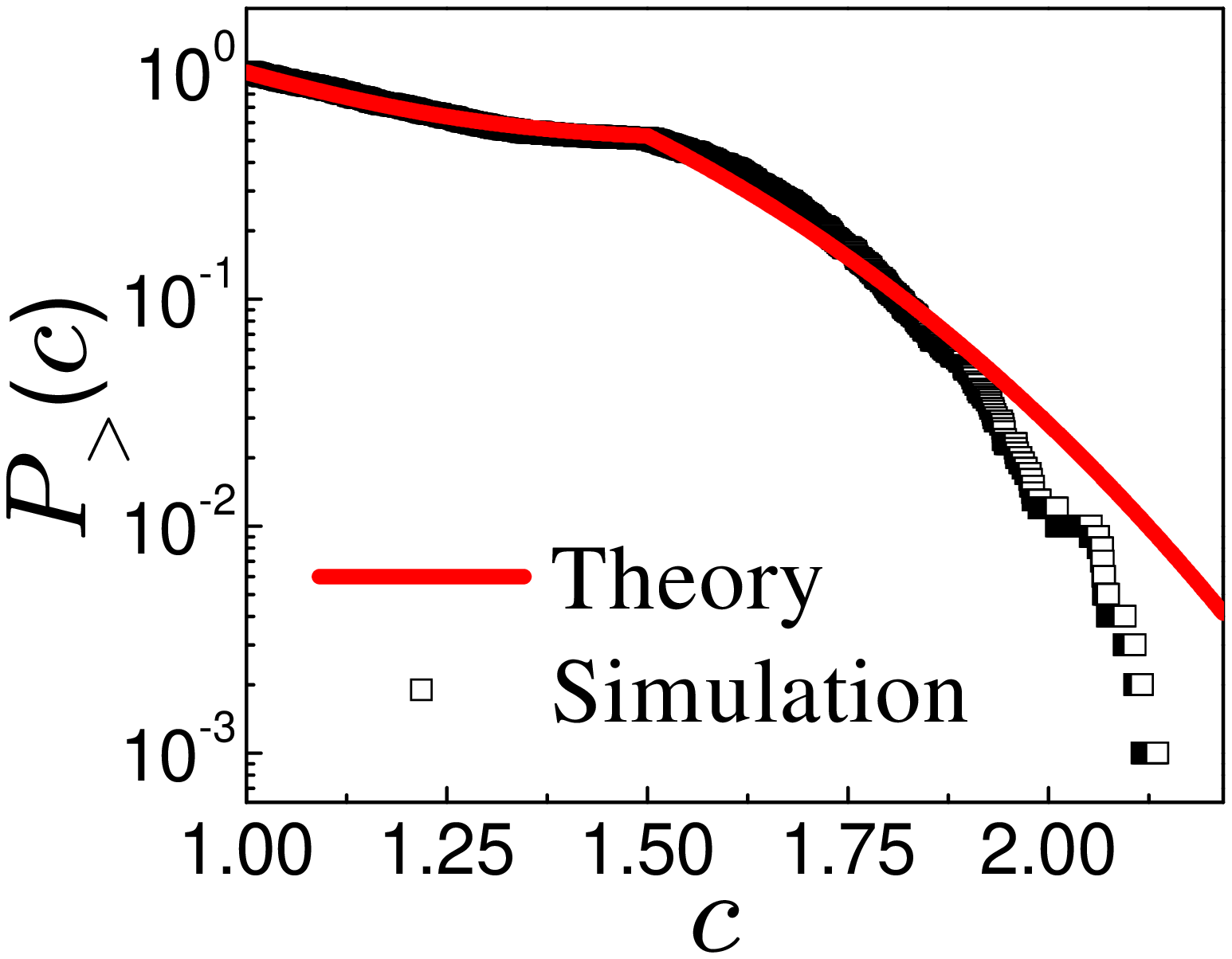}}
        \caption{Comparison between theory and numerical simulations
          for (a) $z_{ij}$, Eq. (\ref{zs}), and (b) $P_>(c)$,
          Eq. (\ref{pc}).}
    \label{fig:zandP}
\end{figure}

Figure ~\ref{fig:zandP}a, compares this solution to numerical
simulations of Hertz packings jammed at RCP \cite{swm} for
$R_1/R_2=1.4$ and $z=6$. We find that the formulae are very accurate
for size ratios below 1.5 and present small deviations up to size
ratio 2. The results are also in agreement with \cite{doods,zamponi}.

{\bf The Voronoi cell.--} The common way to divide a system into
volumes associated with each particle is the Voronoi tessellation. The
Voronoi cell for monodisperse particles \cite{swm} is composed by all
the points nearest to the center of the ball than to any other ball.
This definition has been extended in \cite{troadec} to the case of
polydisperse systems and non-spherical particles: instead of
considering the classical Voronoi polyhedron defined by the center of
the particle, one should consider all the points which are closer to
the surface of a given particle. Such a construction is called the
{\it Voronoi S region} and tiles a system of nonspherical convex
particles and polydisperse systems as can be seen in
Fig. ~\ref{voro}b.  Following this approach we calculate
the average volume of a polydisperse Voronoi cell, denoted $W$. The
volume fraction is given by $\phi=\frac{V_g}{W}$, where $V_g=\frac{4
  \pi}{3}(x R_1^3+(1-x) R_2^3)$ is the mean volume of a ball.  We
first find the analytical formula of the Voronoi S region. The
boundary of the Voronoi cell in the direction $\hat{s}$ of a $i$ ball
next to a $j$ ball at position $r_{ij}$ is (Fig.~\ref{voro}b):

\begin{equation}
  l_{ij}(\hat{s})=\frac{1}{2}\frac{r_{ij}^2-(R_i-R_j)^2}
{r_{ij}(\hat{r}_{ij}\cdot\hat{s})-(R_i-R_j)},
\end{equation}
where $\hat{s}$ and $\hat{r}_{ij}$ are unitary.
Thus, the boundary of a Voronoi cell of a ball $i$ in
the direction $\hat{s}$ is the minimum of $l_{ij}(\hat{s})$ over
all the particles $j$ for any $l_{ij}(\hat{s})>0$. This leads to

\begin{equation}
  l_i(\hat{s})=\frac{1}{2}\min_{\hat{s}.\hat{r}_{ij}>\frac{R_i-R_j}{r_{ij}}}{
\frac{r_{ij}^2-(R_i-R_j)^2}{r_{ij}(\hat{r}_{ij}\cdot\hat{s})-(R_i-R_j)}}.
\end{equation}

The volume of the cell of the ball $i$ is then given by
$W_i=\frac{1}{3} \oint l_i(\hat{s})^3 ds$. We define the orientational
Voronoi volume, $W_i^s$, along the direction $\hat{s}$ by $W_i\equiv
\frac{1}{4\pi}\oint W_i^s ds=\langle W_i^s \rangle_s$. This leads to

\begin{equation}
  W_i^s=\frac{\pi}{6}\min_{\hat{r_{ij}}.\hat{s}>\frac{R_i-R_j}{r_{ij}}}{\Big(\frac{r_{ij}^2-(R_i-R_j)^2}{r_{ij}(\hat{r_{ij}}.\hat{s})-(R_i-R_j)}\Big)^3}.
\label{voronoi}
\end{equation}
This definition leads to the results of \cite{swm} when $R_1=R_2$.
Since the system is isotropic, the mean Voronoi volume can be
calculated as:

\begin{equation}
  W\equiv\langle \langle W_i^s \rangle_s \rangle_i=\langle \langle W_i^s
  \rangle_i \rangle_s=\langle W_i^s \rangle_i.
\end{equation}

{\bf Calculation of the mean Voronoi volume.---} Having calculated the
Voronoi cell exactly in Eq. (\ref{voronoi}), we now proceed to develop
a probability theory of volume fluctuations in the spirit of the
quasiparticle approximation used in \cite{swm} to obtain the mean
Voronoi volume. For a given ball $i$, the calculation of $W^s_i$
reduces to finding the ball $j^*$ that minimizes $l_{ij}(\hat{s})$. We
call $j^*$ the Voronoi ball for the ball $i$. We consider
$r\equiv r_{ij^*}$, $\cos \theta \equiv \hat{s} \cdot \hat{r}_{ij^*}$
and $c\equiv 2 l_{ij^*}$. We have $c=\frac{r^2-(R_i-R_j)^2}{r
  \cos\theta-(R_i-R_j)} $.  Therefore, we just need to compute the
inverse cumulative distribution function, denoted $P_>(c)$, to find all
the balls $j$ with $l_{ij}>\frac{c}{2}$, and thus not contributing to
the Voronoi volume of the ball $i$. The average Voronoi volume is then
given by the expression

\begin{equation}
  W = \frac{\pi}{6} \langle c^3 \rangle
  = -\frac{\pi}{6}\int^{\infty}_{0}c^3 \frac{d
    P_>(c)}{dc}dc=\frac{\pi}{2}\int^{\infty}_{0}c^2 P_>(c)dc. \label{ww}
\end{equation}

We calculate the mean Voronoi volume for the balls of radius $R_1$ and
$R_2$ separately and then average them. We denote $P_>^1(c)$ and
$P_>^2(c)$ the inverse cumulative distributions
respectively, and
$W = x \frac{\pi}{2}\int^{\infty}_{0}c^2
P_>^1(c)dc+(1-x)\frac{\pi}{2}\int^{\infty}_{0}c^2 P_>^2(c)dc,$
and therefore,
\begin{equation}
P_>(c)=x P_>^1(c)+(1-x) P_>^2(c). \label{p_c}
\end{equation}

{\bf Calculation of $P_>(c)$.---} There are three salient steps in the
calculation of $P_>(c)$: {\it (i)} The separation of $P_>(c)$
following Eq. ~(\ref{p_c}).  {\it (ii)} The separation of each term
$P_>^i(c)$, $i=1, 2$, into two contributions: a term taking into
account the contact spheres, $P_>^{i C}(c)$, and a bulk term, $P_>^{i
  B}(c)$. The contact term clearly depends on $z_{ij}$. The bulk term
averages over all spatial
correlations of non-contact particles and, without significant loss of accuracy
as shown below, we assume that it only depends on the
average value of $W$. In principle, it is possible to use a more realistic
form for this term, but this would render the problem practically unsolvable.  {\it (iii)} The separation of $P_>^{i C}(c)$ and
$P_>^{i B}(c)$ into two terms $P_>^{ij C}(c)$ and $P_>^{ij B}(c)$,
$j=1, 2$, for each species.

An assumption of the theory (to be tested a posteriori with computer
simulations) is that all of these terms are independent. Thus

\begin{eqnarray}
  P_>(c) &=& x P_>^{11C}(c)P_>^{12C}(c)P_>^{11B}(c)P_>^{12B}(c) +
\label{pc}
  \\
  &+& (1-x)P_>^{21C}(c)P_>^{22C}(c)P_>^{21B}(c)P_>^{22B}(c).   \nonumber
\end{eqnarray}

Also, we work in the limit of large number of particles leading to
Boltzmann-like exponential distributions for each $P_>^{ijC}$ and
$P_>^{ijB}$ \cite{swm}. Four important quantities are then defined:
{\it (i)} $V^*_{ij}(c)$ and {\it (ii)} $S^*_{ij}(c)$: the excluded
volume and surface on the ball, respectively, where no center of a
ball $j$ can be located for a given ball $i$ and for a given $c$.
{\it (iii)} $\rho_j$: the mean number of balls $j$ by the left free
volume.  {\it (iv)} $\rho^s_{ij}$: the mean number of balls $j$ by
the left free surface on a ball $i$.  These considerations lead to:
\begin{eqnarray}
  P_>^{ijB}(c)=\exp\Big(-\rho_j V^*_{ij}(c)\Big),
\label{pij} \\
P_>^{ijC}(c)=\exp\Big(-\rho^s_{ij} S^*_{ij}(c)\Big). \nonumber
\end{eqnarray}

Next, we calculate these four quantities. To simplify we denote
$l\equiv R_i+R_j$, $k\equiv R_i-R_j$ and $\Theta$ the
step-function. We obtain:

\begin{eqnarray}
  S^{*}_{ij}(c)& =& \int \Theta(c-\frac{l^2-k^2}{l \cos\theta-k)} ds =
  2 \pi(1-\frac{l^2-k^2-kc}{lc}), \nonumber \\
  V^*_{ij}(c)&=& \int \Theta(c-\frac{r^2-k^2}{r \cos\theta-k}) dr^3 = 2\pi(-\frac{1}{4c}((c+k)^4-l^4) + \nonumber \\
  && +\frac{1}{3}((c+k)^3-l^3)
  + (\frac{k^2}{2c}+\frac{k}{2})((c+k)^2-l^2)),\nonumber\\
  \rho_j(W)&=&\frac{x_j}{W-V_g},
\label{sstar}
\end{eqnarray}
where $x_1=x$ and $x_2=(1-x)$. The fourth quantity, $\rho^s_{ij}$, is
the most difficult to calculate. In terms of the occupied areas
Eqs. (\ref{occ}) we have
$\rho^s_{ij}=\frac{z_{ij}}{4\pi-z_{i1}S^{occ}_{i1}-z_{i2}S^{occ}_{i2}}$.
However, for the contact terms, the analogy with the Boltzmann-like
exponential distribution of volumes is far from being exact. This is
because this form is based on the large number limit which in the case
of contacting balls is no more than around 6. Therefore, the
exponential distribution is used as an Ansatz with $\rho^s_{ij}$ a
variational parameter as in \cite{swm}. We denote $\langle S^{\rm
  free}_{ij} \rangle$ the mean solid angle of the gaps left between
the $j$ contacting neighbors of a $i$ ball (Fig. ~\ref{voro}a). We
obtain:

\begin{eqnarray}
\langle S^{\rm free}_{ij}\rangle &=& \int^{\infty}_{0}S^*_{ij}(c)\frac{d(1-P_>(c))}{dc}dc \\
&\approx & \rho^s_{ij}\int^{2\pi}_{0}S^*_{ij}\exp(-\rho^s_{ij}S^*_{ij})dS^*_{ij}=\frac{1}{\rho^s_{ij}}. \nonumber
\end{eqnarray}

Then, we perform numerical simulations to find $\langle S^{free}_{ij}
\rangle$. We randomly generate balls of radius $R_i$ and $R_j$ with
the proportion $z_{i1}/z_i$ and $z_{i2}/z_i$ respectively around a
ball of radius $R_i$ and then evaluate the mean free surface $\langle
S^{free}_{ij}\rangle$ and its inverse to obtain $\rho^s_{ij}$.  We
find

\begin{eqnarray}
  \rho^s_{ij}(z_{ij}) &=&
  \frac{z_{ij}}{2\pi}(1-\frac{z_{i1}}{z_{i}}\frac{S^{occ}_{i1}}{2\pi}-\frac{z_{i2}}{z_{i}}\frac{S^{occ}_{i2}}{2\pi}) =
\label{rhos}
\\
& = &
  \frac{z_{ij}}{2\pi}(\frac{z_{i1}}{z_{i}}\sqrt{1-(\frac{R_1}{R_i+R_1})^2}+\frac{z_{i2}}{z_{i}}\sqrt{1-(\frac{R_2}{R_i+R_2})^2})),
  \nonumber
\end{eqnarray}
which is a generalization of the results of \cite{swm} to polydisperse
systems.

Using Eqs. (\ref{pij}), (\ref{sstar}) and (\ref{rhos}) into
(\ref{pc}), $P_>(c)$ can be calculated by solving the equations
numerically.  Figure \ref{fig:zandP}b depicts the comparison of
the theoretical results of the probability of Voronoi volumes
$P_>(c)$ with computer generated Hertzian packings with $z=6$ for
$x=0.5$ at RCP. The calculated distribution is quite accurate for
most of the range except for small deviations at high values of
$c$, which however, do not affect much the value of the average
$\langle c^3 \rangle$ in Eq. (\ref{ww}). This shows that
our mean-field approximation already captures the main contribution of
the probability distribution function $P_>(c)$.

\begin{figure}[t] \centering
(a)
\includegraphics[width=0.25\textwidth]{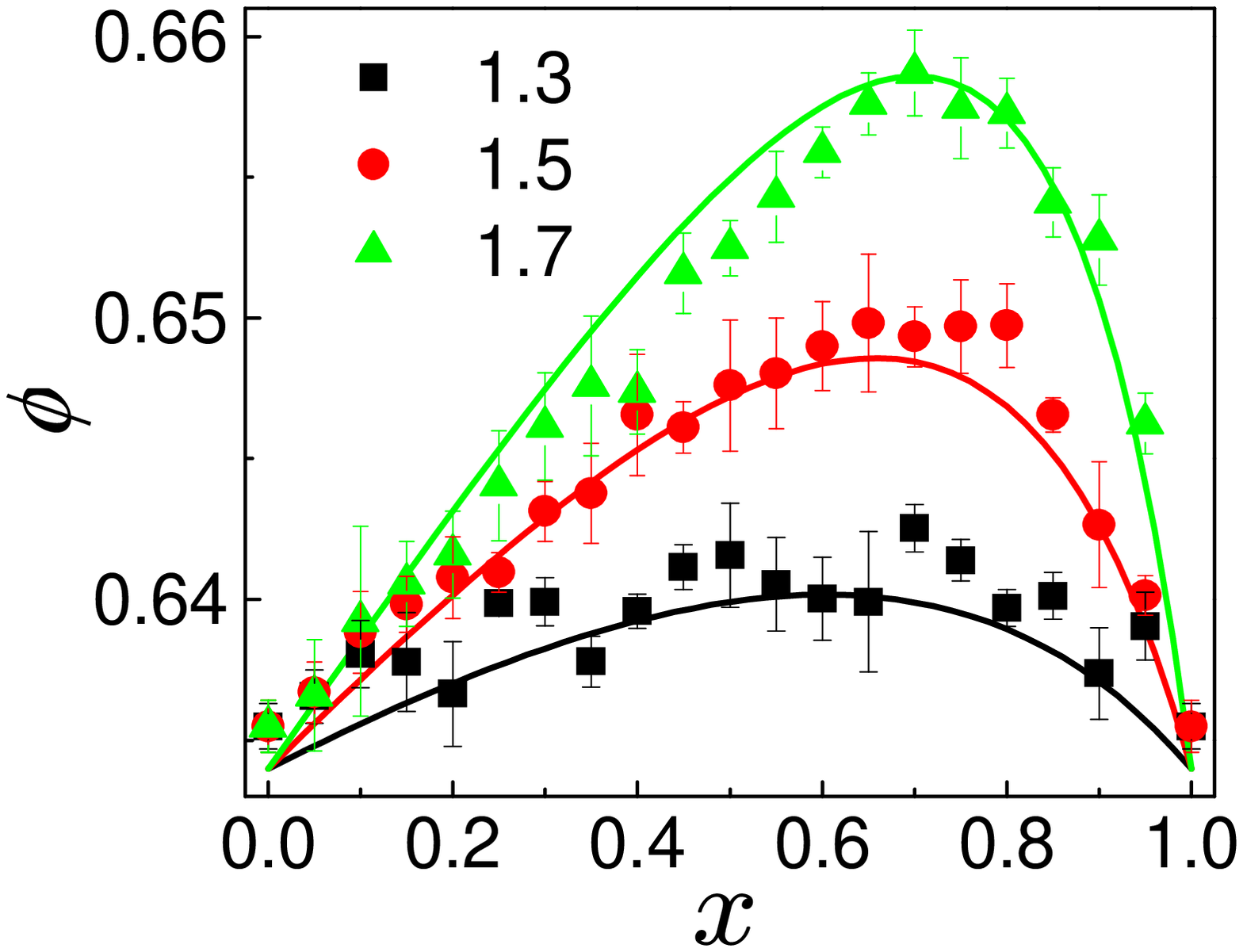}
(b)
\includegraphics[width=0.25\textwidth]{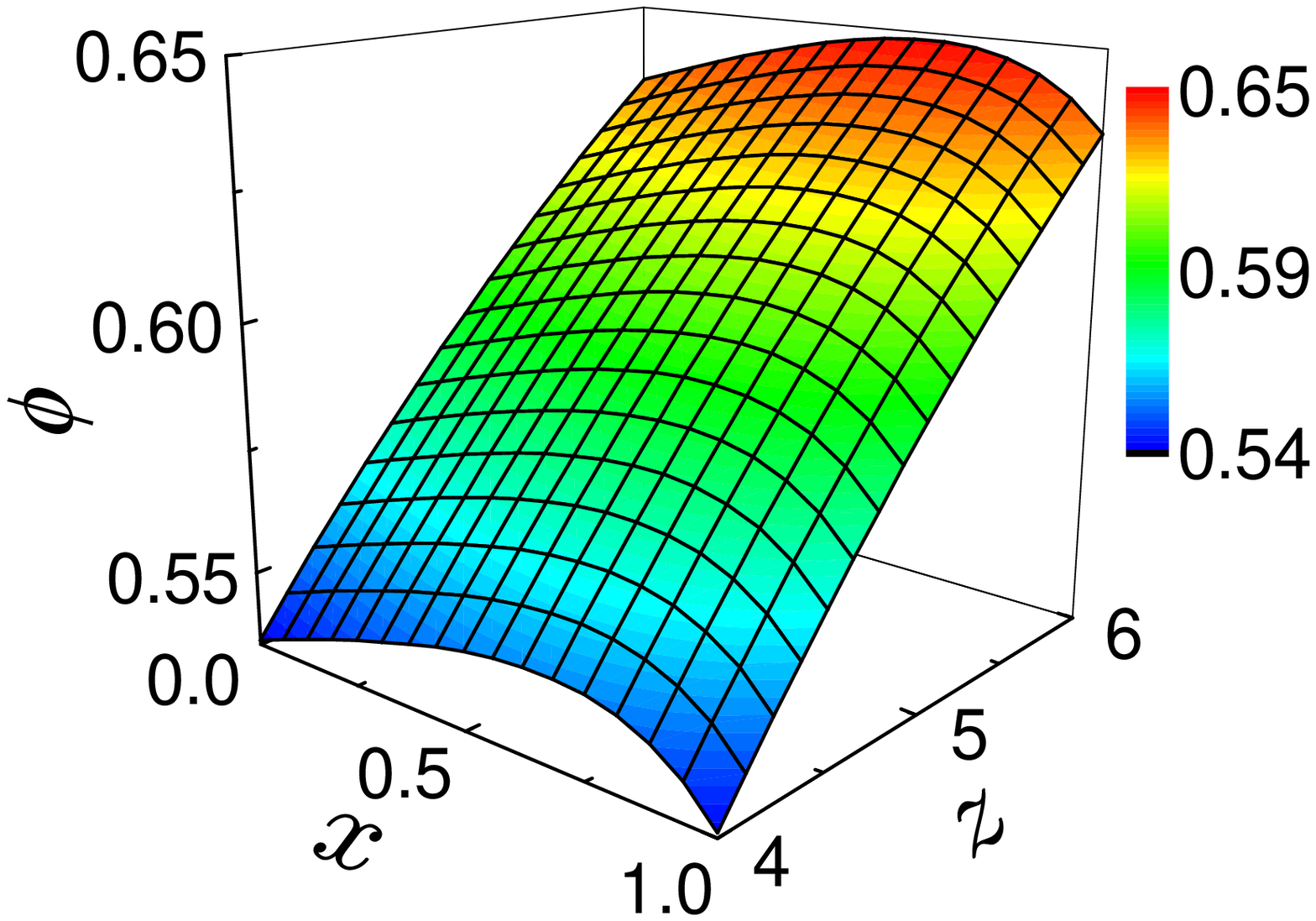}
\caption{(a) Comparison between theory and numerical simulations of
  Hertzian packings at RCP, i.e. $z=6$ versus $x$ for different values
  of $R_1/R_2$ as indicated. Error bars are std over 10 realizations
  of the packings with 10,000 balls. (b) Three dimensional surface
  plot of $\phi$ as a function of $z$ and $x$ for $R_1/R_2=1.5$.
  The numerical results at RCP and RLP are provided in Supplementary Information A.}
\label{fig:jamcompare}
\end{figure}

{\bf Calculation of $W$.---} The above considerations lead to a final
form to calculate $ W$ using Eq. (\ref{pc}) into (\ref{ww}):
\begin{equation}
  W=\frac{\pi}{2}\sum_{i}x_i \int^{\infty}_{0}c^2\exp\Big(\sum_{j}(-\rho^s_{ij}
  S^{*}_{ij}(c)-\rho_{j}(W) V^{*}_{ij}(c))\Big)dc. \label{final}
\end{equation}
Notice that $\rho_j(W)$ depends on $W$, Eq. (\ref{sstar}), and
$\rho^s_{ij}(z_{ij})$ depends on the $z_{ij}$, Eq. (\ref{rhos}), which
in turn depends on the concentration $x$ and $z$ through
Eq. (\ref{zs}). Therefore Eq. (\ref{final}) is a self-consistent
equation to obtain $W(z,x)$, for a given $R_1/R_2$.  A numerical
integration of Eq. (\ref{final}) is performed versus $x$ for a given
value of $z$.  By considering the isostatic limits of $z=6$ and $z=4$
we predict the limits of RCP and RLP at zero friction and infinite
friction between the spheres, respectively \cite{swm}. The results for
the volume fraction at RCP versus $x$ are depicted in
Fig. ~\ref{fig:jamcompare}a which also compares the results to
numerically generated packings of Hertz spheres \cite{swm}.  We see a
very good agreement indicating that the theory captures well the
behavior of polydisperse packings and that the approximations used are
reasonable. For size ratios larger than 2 deviations are found
indicating the limit of validity of the approach.  For any other value
of interparticle friction between 0 and $\infty$, the density is
obtained by setting $z$ between the limiting isostatic values of 6 and
4, respectively. The resulting volume fraction is shown in
Fig. \ref{fig:jamcompare}b.  The result for RLP for a given $x$ is a
new prediction as this problem has not been investigated before. Our
results promote new experiments to test the RLP limit of polydisperse
systems shown in Fig. \ref{fig:jamcompare}b.

The formalism can be extended to consider any distribution of sphere
radius.  The main modification is that all the sums over the radius
are replaced by integrations over the desired distribution of
radius $P(R)$ (the binary case corresponds to two delta-functions at
$R_1$ and $R_2$).

And all the quantities are calculated for balls of internal radius
$r$ and external $R$ and $x$ and $(1-x)$ are replaced by $P(r)dr$ and
$P(R)dR$, respectively, including the coordinations (see Supplementary Information B). This result
allows to explore the space of radius distributions in search of the
optimal packings for a given polydispersivity. This analysis could be
of industrial interest if one wishes to optimize the packing fraction
by introducing different species in the mixture.

We calculate the volume fraction for several distributions $P(R)$
constraint by ball radius in the range [1,2], in search of the optimal
packing. We calculate the volume fraction
for various $P(R)$ ranging from uniform to two-peaked Gaussian
distributions of varied widths. We find that the more small balls we
add the better the packing until a certain point where the volume
fraction starts to decrease. This maximum can be rationalized assuming
that the small balls always fill the gaps between the large ones as
long as there are enough large balls.  Further extensions of the
theory to any dimension can be performed by replacing $3$ by d in
Eq. (\ref{voronoi}) and developing a theory of volume fluctuation in
d-dimensions. We notice that many of the approximations employed in 3d
may become exact for large d, thus we expect improved results in the
mean field limit of infinite dimensions.

The method allows to treat more difficult problems. For instance, the
prediction of the volume fraction of a jammed system of non-spherical
particles is a long-standing problem. Theoretical predictions of
Onsager \cite{onsager}
are valid for large aspect ratios, like elongated rods.  Experiments
however, find interesting new physics for small aspect ratios. In this
respect, the present polydisperse theory could be mapped to the
problem of ellipsoids, spherocylinders or rods. A Voronoi cell needs
to be calculated as a function of the angles defining the orientation
of the non-spherical particles in analogy of the calculation between
two particles of different radii. The integration over $P(r) dr$ in
Eq. (\ref{final}) is then replaced by integration over weighted
orientational angles. The above analysis can also be extended to dimensions beyond three \cite{Meel}.
Although many of the appproximations should work better in higher dimensions, some of the
hypotheses (for example, the contact term ansatz) need to be reassessed.
Thus, higher dimensions studies cannot be addressed as trivial extensions and
need to be handled with care.

In summary, a theoretical framework is presented that predicts the RLP
and RCP limits of a system of polydisperse spheres and brings together
distinct results into a common theoretical framework. The formalism
has the potential to solve other problems in condensed matter physics
such as the mixing and phase behavior of systems of hard particles of
different shapes and size.

\newpage
\strut
\newpage

\begin{widetext}

\appendix

\section{Table of volume fraction values predicted by the theory.}

\begin{table}[H]
	\centering
		\begin{tabular}{|c|c|c|c|c|c|c|c|c|c|c|}   \hline
    \hline
    \backslashbox {~~~~$R_2/R_1$}{x~~~~} & 0 & 0.1 & 0.2 & 0.3 & 0.4 & 0.5 & 0.6 & 0.7 & 0.8 & 0.9 \\ 
    \hline
    1.3 & 0.5359 & 0.5376 & 0.5392 & 0.5405 & 0.5416 & 0.5423 & 0.5426 & 0.5423 & 0.5413 & 0.5393 \\ 
    1.5 & 0.5359 & 0.5393 & 0.5426 & 0.5456 & 0.5482 & 0.5503 & 0.5516 & 0.5517 & 0.5499 & 0.5453 \\
    1.7 & 0.5359 & 0.5409 & 0.5457 & 0.5506 & 0.5550 & 0.5588 & 0.5616 & 0.5628 & 0.5611 & 0.5359\\
    \hline
    \hline
  \end{tabular}
	\caption{Volume fraction for $z=4$, RLP.}
	\label{tab:cap1}
\end{table}

\begin{table}[H]
	\centering
		  \begin{tabular}{|c|c|c|c|c|c|c|c|c|c|c|}
    \hline
    \hline
    \backslashbox {~~~~$R_2/R_1$}{x~~~~} & 0 & 0.1 & 0.2 & 0.3 & 0.4 & 0.5 & 0.6 & 0.7 & 0.8 & 0.9 \\ 
    \hline
    1.3 & 0.6340 & 0.6356 & 0.6370 & 0.6382 & 0.6392 & 0.6399 & 0.6402 & 0.6399 & 0.6389 & 0.6340 \\ 
    1.5 & 0.6340 & 0.6372 & 0.6401 & 0.6423 & 0.6453 & 0.6472 & 0.6484 & 0.6484 & 0.6468 & 0.6426 \\
    1.7 & 0.6340 & 0.6386 & 0.6431 & 0.6475 & 0.6515 & 0.6549 & 0.6575 & 0.6586 & 0.6571 & 0.6506\\
    \hline
    \hline
  \end{tabular}
	\caption{Volume fraction for $z=6$, RCP.}
	\label{tab:cap}
\end{table}

\section{Distribution of sphere radius.}

According to the theory, the average Voronoi volume for a packing with a distribution of radius $P(r)$, is given by the following self-consistent equation :

\begin{equation}
  W=\frac{\pi}{2}~\int dr P(r)\!\! \! \int^{\infty}_{0}\!\!\!\! \!\!c^2\exp\Big(\int dR (-\rho^s_{rR}
  S^{*}_{rR}(c)-\rho_{R}(W) V^{*}_{rR}(c))\Big)dc,  \label{final2}
\end{equation}

where the different quantities are calculated as follow:
$$S^{*}_{rR}(c) = \int \Theta(c-\frac{l^2-k^2}{l \cos\theta-k}) ds = 2\pi(1-\frac{l^2-k^2+kc}{lc}),$$
$$V^*_{rR}(c)= \int \Theta(c-\frac{r^2-k^2}{r \cos\theta-k}) dr^3= 2\pi(-\frac{1}{4c}((c-k)^4-l^4) +\frac{1}{3}((c-k)^3-l^3)  + (\frac{k^2}{2c}-\frac{k}{2})((c-k)^2-l^2)),$$
To simplify we denoted $l= r+R$, $k= r-R$ and $\Theta$ the step-function.
$$\rho_{R}(W)=\frac{P(R)}{W-V_g},$$
$$\rho^s_{rR}=\frac{z_{rR}}{2\pi}\int^{\infty}_{0}\frac{z_{rr'}}{z_r}\sqrt{1-(\frac{r'}{r+r'})^2})dr',$$
$$z_r=z\frac{A}{\langle S^{occ}_r \rangle},$$
$$\langle S^{occ}_r \rangle=\int^{\infty}_{0}2\pi (1-\sqrt{1-(\frac{r'}{r+r'})^2})P(r')dr',$$
$$A^{-1} = \int^{\infty}_{0}\frac{P(r')}{\langle S^{occ}_{r'} \rangle} dr',$$
$$z_{rr'}=\frac{z_r z_{r'} P(r')}{z}.$$

\end{widetext}


\begin{thebibliography}{99}



\bibitem{coniglio} A. Coniglio, A. Fierro, H. J. Herrmann,
  M. Nicodemi, eds, \emph{Unifying concepts in granular media and
    glasses} (Elsevier, Amsterdam, 2004).


\bibitem{bernal} J. D. Bernal, J. Mason,
  Nature {\bf{188}}, 910 (1960).



\bibitem{doods} J. Dodds, Nature {\bf 256}, 187 (1975);
K. de Lange Kristiansen, A. Wouterse, A. Philipse, Physica A
  {\bf 358}, 249 (2005); M. Clusel, E. I. Corwin, A. O. N. Siemens,
  J. Bruji\'c, Nature {\bf 460}, 611 (2009).

\bibitem{zamponi} I. Biazzo, F. Caltagirone, G. Parisi, F. Zamponi,
  Phys. Rev. Lett. {\bf 102}, 195701 (2009).











\bibitem{swm} C. Song, P. Wang, H. A. Makse, Nature {\bf 453}, 629
  (2008); C. Briscoe, C. Song, P. Wang, H. A. Makse,
  Phys. Rev. Lett. {\bf 101}, 188001 (2008).


\bibitem{onsager} L. Onsager, Ann. N. Y. Acad. Sci. {\bf 51}, 627
  (1949).





\bibitem{troadec} V. A. Luchnikov, N. N. Medvedev, L. Oger,
  J.-P. Troadec,
  Phys. Rev. E {\bf 59}, 7205 (1999).

\bibitem{Meel}  J. A. van Meel, B. Charbonneau, A. Fortini, P. Charbonneau Phys. Rev. E, 80,061110, (2009).

\end{thebibliography}
\end{document}